# Hertz-level Measurement of the $^{40}$Ca$^+$ 4s $^2S_{1/2}$–3d $^2D_{5/2}$ Clock Transition Frequency With Respect to the SI Second through GPS

Y. Huang,[1,2] J. Cao,[1,2] P. Liu,[1,2] K. Liang,[3] B. Ou,[1,2,†] H. Guan,[1,2] X. Huang,[1,2] T. Li[3] and K. Gao[1,2,*]

[1]*State Key Laboratory of Magnetic Resonance and Atomic and Molecular Physics*

[2]*Key Laboratory of Atomic Frequency Standards,*

*Wuhan Institute of Physics and Mathematics, Chinese Academy of Sciences, Wuhan 430071, China*

[3]*National Institute of Metrology, Beijing 100013, China*

We report a frequency measurement of the clock transition of a single $^{40}$Ca$^+$ ion trapped and laser cooled in a miniature ring Paul trap with $10^{-15}$ level uncertainty. In the measurement, we used an optical frequency comb referenced to a Hydrogen maser, which was calibrated to the SI second through the Global Positioning System (GPS). Two rounds of measurements were taken in May and June 2011, respectively. The frequency was measured to be 411 042 129 776 393.0(1.6) Hz with a fractional uncertainty of $3.9 \times 10^{-15}$ in a total averaging time of $> 2 \times 10^6$ s within 32 days.

Optical frequency standards have been developed rapidly in recent years thanks to the development of the cold atoms, the optical frequency comb technology [1, 2] and the ultra-narrow-linewidth lasers. Optical frequency standards are expected to replace the Cs primary microwave standard as in the definition of the SI second in the near future. The measurement of the absolute optical frequency for the clock transition is an important step in the development of the optical frequency standards based on single ions and ultracold neutral atoms. The clock transition frequency had been measured using the optical frequency comb referenced to the Cs fountain and using GPS as a link to the SI second without a primary standard for direct calibration. The clock transition frequency had been measured referenced to the Cs fountain at uncertainties on the order of $10^{-15}$ or even smaller with $^{87}$Sr, $^{88}$Sr, $^{171}$Yb$^+$, $^{27}$Al$^+$, $^{40}$Ca$^+$ and $^{199}$Hg$^+$ [3-9]. A Sr lattice clock frequency had been measured referenced to GPS as a link to the SI second by National Metrology Institute of Japan/National Institute of Advanced Industrial Science and Technology (NMIJ/AIST) at the uncertainties on the order of $10^{-14}$ level [10]. With the developments of the precise measurements on the optical frequency standard either based on single trapped ion or ultracold neutral atoms, lasers (optical local oscillators) frequency stabilized to the transitions of atoms/ions were recommended by the International Committee for Weights and Measures (CIPM) as secondary representations of the SI second, contributing to International Atomic Time (TAI) [11].

We are developing an optical frequency standard based on a single $^{40}$Ca$^+$ ion, which has been proposed as one of the candidates of the future frequency standards for the next definition of the SI second[11]. The 4s $^2S_{1/2}$–3d $^2D_{5/2}$ clock transition of $^{40}$Ca$^+$ at 729 nm has a natural linewidth of 0.2 Hz[12], which has good potential accuracy and low systematic shifts. The optical frequency standard based on $^{40}$Ca$^+$ is also being developed by the Quantum Optics and Spectroscopy Group in Innsbruck and the National Institute of Information and Communications in Japan (NICT). In Innsbruck, the frequency was measured to be 411 042 129 776 393.2(1.0) Hz with a fractional uncertainty of $2.4 \times 10^{-15}$, which was referenced to the transportable Cs atomic fountain clock of LNE-SYRTE [8]; in NICT, the frequency was measured to be 411 042 129 776 385($\pm$18) Hz with a fractional uncertainty of $10^{-14}$ level, which was referenced to the Cs atomic fountain clock [13].

In this paper, we report on the first measurements of the $^{40}$Ca$^+$ 4s $^2S_{1/2}$–3d $^2D_{5/2}$ transition frequency with an uncertainty level of $10^{-15}$ using an optical frequency comb referenced to a Hydrogen maser, which was calibrated through the Global Positioning System (GPS) as a link to the SI second. To approach a measurement of $10^{-15}$ level accuracy, we performed the measurement for a very long time ($7 \times 10^5$ s in May and $1.5 \times 10^6$ s in June) to reduce the statistical error and the frequency transfer error. At the same time, the systematic error had been reduced to be smaller than the previous work [14] by achieving the lower ion temperature and increasing the measurement precision on the electric quadrupole shift.

Figure 1 is a schematic diagram of our experimental setup. Full details of the laser cooling, trapping detecting and probing system including the locking scheme of probe laser to the ion transitions used in this work were reported in previous works [14-17]. Briefly, a single $^{40}$Ca$^+$ ion was trapped and cooled in a miniature Paul ring trap. The trap had an endcap-to-center distance of $z_0 \approx 0.6$mm with a center-to-ring electrode distance $r_0 \approx 0.8$ mm. Two electrodes perpendicular to each other were set in the ring plane to compensate the ion's excess micromotion. A trapping rf of ~580 $V_{p-p}$ was applied to the ring at a frequency of 9.8 MHz. The excess micromotion was nulled by applying different dc bias voltages on the endcap electrodes and the compensation electrodes. The 729 nm laser was locked to the six chosen Zeeman components to cancel the linear Zeeman shift and the electric quadrupole shift. After that, the measurements of the clock transition frequency, the stability of the clock, and the evaluation of the systematic shifts were taken.

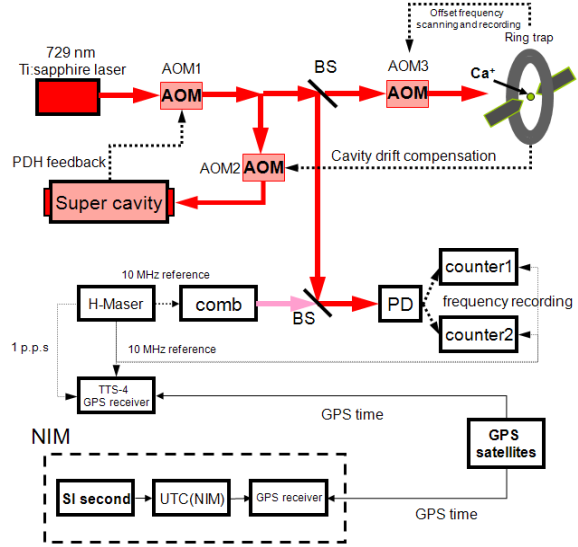

Fig. 1. (color online) Overall schematic of the frequency measurement experimental setup. AOM: acousto-optic modulator; HWP: half wave plate; PD: photo detector; GPS: global positioning system, NIM: National Institute of Metrology of China. AOM1 is for the locking of the laser to the super cavity as the fast loop feedback of the PDH method, AOM2 is for the compensation of the cavity drift, AOM3 is for the offset frequency scanning and recording.

The probe laser at 729 nm was a commercial Ti: sapphire laser (MBR-110, Coherent), which was locked to a temperature-controlled high-finesse Fabry-Perot cavity (Zerodur material) supported on an active isolated platform (TS-140, Table stable) using Pound-Drever-Hall technique [18]. A linewidth of ~13 Hz was measured from the heterodyne beatnote with a home-made diode laser stabilized to another high-finesse ultra-low-expansion (ULE) cavity. The long-term drift was measured to be ~3 Hz/s. An acousto-optic modulator (AOM) (80 MHz, Brimrose) driven by a sweeping function generator (2023A, IFR) was used to compensate the long term drift. After the compensation, we got a non-linear drift of <0.1 Hz/s. The offset frequency between the probe laser and the clock transition line center of the ion was achieved by an AOM, which shifted the laser frequency to match the transitions. The required AOM frequencies were updated every 40 cycles of pulses, which cost ~1.1 s. By the "four points locking scheme" [19, 20], 3 pairs of the Zeeman transitions were interrogated, and the offset frequency between the probe laser and the clock transition could be obtained every ~13 s. In the mean time, a Ti: sapphire based optical frequency comb (FC 8004, MenloSystems) was used to measuring the 729 nm laser frequency. Both the repetition frequency and the offset frequency of the fs comb were locked to two individual synthesizers, which were referenced to a 10 MHz signal provided by an active Hydrogen maser (CH1-75A) with an isolated splitter and a 60-m-long standard 50 Ohm coaxial cable. Two individual counters were used to measure the beat frequency of the comb laser and the 729 nm laser simultaneously, if the difference of the readings of the two counters were >1 Hz, we believe the measurement was not reliable thus not taken into count. The probe laser frequency was measured every 1 s. With the above two parts of the measurement results, we can do the measurement of the clock transition frequency referenced to the H-maser and calculate the frequency instability comparison of the $^{40}Ca^+$ optical clock vs. the H-maser (Fig. 2). Fig. 2 shows the histogram of the clock transition frequency measurements referenced to the H-maser on the day MJD 55 726, which gives an averaged value of 411 042 129 776 490.7 Hz. The histogram follows a normal distribution and the standard deviation of the mean is 3.5 Hz. The longest continuous measurement was up to >50 h. As shown in

Fig. 2, the Allan deviation for the ion transition vs. Hydrogen maser comparison reaches the $10^{-15}$ level after >2 000 s of averaging time, which could be limited by the stability of the H-maser and the stability of the frequency transfer.

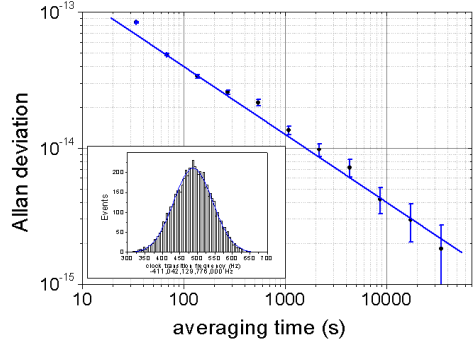

Fig. 2. (color online) The Allan deviation for the $^{40}Ca^+$ optical clock vs. the Hydrogen maser, the blue line represents $4\times10^{-13}\times\tau^{-1/2}$. Inset: Histogram of the frequency measurements of the clock transition calculated from the offset frequency and the comb measurements referenced to the H-Maser with a Gaussian fitting (blue line).

Frequency measurements similar as above were taken in 32 individual days, separated into two parts, one in May 2011 with 15 continuous days and the other in June 2011 with 17 continuous days (Fig.3). Each filled circles in Fig. 3 represents a mean value of the measurement result based directly on the Hydrogen maser. The error bars are given by the standard deviation of the mean. The former 15 days of measurements gives a weighted averaged frequency of 411 042 129 776 489.7(0.9) Hz and the later 17 days of measurements gives a weighted averaged frequency of 411 042 129 776 489.1(0.4) Hz. The difference between two rounds of measurements mainly was: in the first round, the average time the laser locked to the ion transition in one day was ~12 h; for the second round, the laser was locked to the transition for >22 h in a day, which means the clock was running for >90% of the time, thus the statistical error for one day of the averaging time was expected to be smaller.

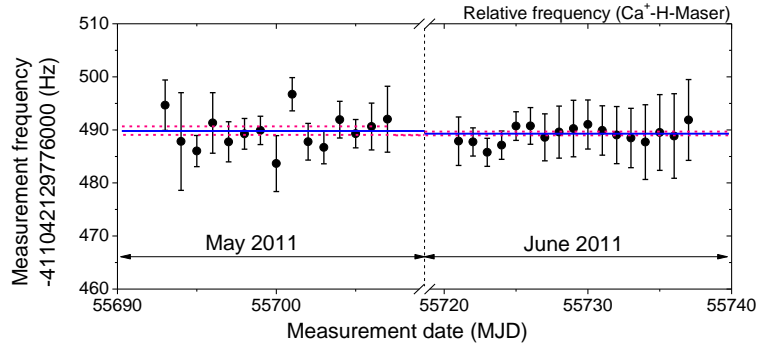

Fig. 3. Frequency measurement of the $4s\ ^2S_{1/2}–3d\ ^2D_{5/2}$ transition of a laser cooled trapped single $^{40}Ca^+$ ion referenced to the Hydrogen maser. Data shown in this figure do not include systematic corrections.

To get the final absolute frequency measurement of the clock transition, systematic shifts and the calibration of the reference must be considered and applied to the above averaged frequency. There are a variety of potential sources of systematic shift which might be associated with the quadrupole 729 nm $4s\ ^2S_{1/2}–3d\ ^2D_{5/2}$ clock transition in a laser cooled trapped $^{40}Ca^+$ ion. The detailed study on the evaluation of the systematic shifts can be found in Ref. [14]. In this paper, systematic shifts are reevaluated. Some improvements were made on the evaluation of the systematic error as follows. After the re-minimization of the micromotion, the ion temperature is estimated to be lower at 3(3) mK from the intensity of the secular sidebands relative to the carrier (normally about 0.2~0.4) [21]. With the ion temperature estimated, the second order Doppler shift caused by thermal kinetic energy is calculated to be -0.004(0.004) Hz [22]. The probe laser was locked to the six different chosen Zeeman transitions one after another. By averaging the center frequency of the three pairs of the components, we can null the quadrupole shift [19, 22, 23]. By averaging the difference of center frequency for different components, the error is obtained. For the much longer averaging time, the fractional error evaluated is smaller than the previous work [14]. Using the new factor value calculated by Ref. [24], the evaluation of the blackbody radiation shift error is reduced. For our $^{40}Ca^+$, the altitude was measured to be 35.2 (1.0) m, thus the gravitational shift for the clock is calculated to be 1.583(0.045) Hz.

Table I shows the total summary of the frequency shifts considered above. Considering all of them; we get a total shift of 1.95 Hz with a fractional error of $5.0 \times 10^{-16}$ in May 2011 and a total shift of 1.95 Hz with a fractional error of $6.5 \times 10^{-16}$ in June 2011.

TABLE I. The systematic frequency shifts with fractional error in $10^{-16}$

| Effect | Measurements in May | | Measurements in June | |
|---|---|---|---|---|
| | Shift (Hz) | Fractional error ($10^{-16}$) | Shift (Hz) | Fractional error ($10^{-16}$) |
| $2^{nd}$ order Doppler shift due to thermal motion | -0.004 | 0.10 | -0.004 | 0.10 |
| $2^{nd}$ order Doppler shift due to micromotion | -0.02 | 0.49 | -0.02 | 0.49 |
| Stark shift due to thermal motion and micromotion | 0 | 0.04 | 0 | 0.04 |
| ac Stark shift due to 397 nm, 866 nm and 854 nm | 0 | 0.04 | 0 | 0.04 |
| ac Stark shift due to 729 nm | 0.04 | 1.46 | 0.04 | 1.46 |
| Blackbody radiation shift | 0.35 | 0.27 | 0.35 | 0.27 |
| Linear Zeeman shift | 0 | 4.52 | 0 | 6.23 |
| $2^{nd}$ order Zeeman shift | 0 | 0.01 | 0 | 0.01 |
| Electric quadrupole shift | 0 | 0.72 | 0 | 0.51 |
| Gravitational shift | 1.58 | 1.10 | 1.58 | 1.10 |
| Total shift | 1.95 | 5.0 | 1.95 | 6.5 |

For our measurement, the largest frequency correction comes from the calibration of the frequency of the H-maser. To calibrate the frequency of the H-maser, a GPS time and frequency transfer receiver with an antenna (TTS-4, PikTime Systems) was used [25]. The receiver with reference to the 10 MHz and 1 pulse per second (p.p.s) signals from the H-maser received the GPS signals from 6~10 GPS satellites on average and generated and recorded the GPS measurement data. In the mean time, a similar receiver with reference to the 10 MHz and the 1 p.p.s signals of the UTC(NIM) in the National Institute of Metrology (NIM) of China did the similar job. Using the two sets of data from the two institutes, we could calculate the frequency difference of the H-maser from the UTC(NIM). Considering the frequency difference of UTC(NIM) and SI-second, the Hydrogen maser we used for frequency measurement can be calibrated. After the calculation with the GPS precise point positioning (PPP) technique [26], we achieved a frequency transfer uncertainty of $\sim 1 \times 10^{-14}$ with an averaging time of one day. Fig. 4 shows the time difference between the H-maser and the UTC (NIM). The weighted average of the frequency offset between the Hydrogen maser and the UTC(NIM) is then calculated to be -2.3649(0.0337)$\times 10^{-13}$ in May 2011 and -2.3582(0.0083)$\times 10^{-13}$ in June 2011. In the mean while, the frequency difference between the UTC (NIM) and the SI-second comes from the primary standard can be calculated from the data reported on the BIPM website [27] to be 5.5(2.1) $\times 10^{-15}$ in May 2011 and 6.0(2.1) $\times 10^{-15}$ in June 2011. Therefore, we calculated the absolute frequency of the $^{40}Ca^+$ clock transition, which is shown in Table II.

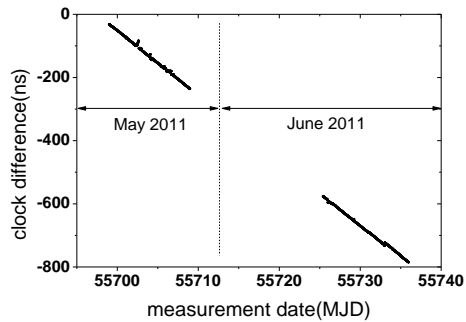

Fig. 4. Time difference between the Hydrogen maser and the UTC (NIM) calculated using PPP technique.

TABLE II. The absolute frequency measurement budget table

| Contributor | Measurements in May | | Measurements in June | |
|---|---|---|---|---|
| | Shift (Hz) | Fractional error ($10^{-15}$) | Shift (Hz) | Fractional error ($10^{-15}$) |
| Systematic shift (Table I) | 1.95 | 0.50 | 1.95 | 0.65 |
| Statistical | 0 | 2.19 | 0 | 0.97 |
| Hydrogen maser reference calibrated with UTC (NIM) | 97.21 | 3.37 | 96.93 | 0.83 |
| UTC (NIM) reference | -2.3 | 2.1 | -2.5 | 2.1 |
| Total | 96.9 | 4.6 | 96.4 | 2.6 |

From Table II, we determined the total correction for the frequency measurement described in Fig. 3 was -96.9 Hz in May 2011 and -96.4 Hz in June 2011. The combined fractional uncertainty of the absolute frequency measurement was $4.6 \times 10^{-15}$ in May 2011 and $2.6 \times 10^{-15}$ in June 2011. The obtained corrected absolute frequency of the $^{40}Ca^+$ $4s$ $^2S_{1/2}$–$3d$ $^2D_{5/2}$ clock transition is 411 042 129 776 393.3(1.9) Hz in May 2011 and 411 042 129 776 392.7(1.1) Hz in June 2011, the two measurements agrees with each other within their uncertainties. The unweighted mean of the above two values gives a final results of 411 042 129 776 393.0(1.6) Hz, the final uncertainty is calculated by considering both the statistical error and the systematical error. The result is in agree with the former measurements [8, 11] by University of Innsbruck and the NICT, also the frequency value recommended by CIPM [10] within their uncertainties. The systematic errors for the clock transition frequency (Table I) considered above are all smaller than $1 \times 10^{-15}$.

The realization of the absolute frequency measurement of the $^{40}Ca^+$ clock transition frequency is a milestone towards a real optical clock and it's important to increase the measurement precision in the precision spectroscopy research area. Hence, the better result could be achieved if we performed the measurement referenced to GPS as a link to the SI second for a longer time. We shall do the comparison of the $^{40}Ca^+$ clocks between Wuhan, Tokyo and Innsbruck, with respect to the SI second through GPS in the following step. Moreover, we suppose the stability of the Hydrogen maser and the transfer cable might cause the problem of the reproducibility that was not good enough and need to be approved. To achieve a smaller uncertainty in the future, we may need to use a more stable reference such as Cs fountain instead of GPS system to achieve better results with smaller measurement uncertainty or to reduce the averaging time achieving the same uncertainty.

We gratefully acknowledge H. Shu, H. Fan, B. Guo, Q. Liu and W. Qu for the early works. We also thank G. Huang for help and working with us. Thank J. Ye, F.-L. Hong, H. Klein, K. Matsubara, M. Kajita, Y. Li and L. Ma for help and fruitful discussions. This work is supported by the National Basic Research Program of China (2005CB724502) and (2012CB821301), the National Natural Science Foundation of China (10874205, 10274093 and 11034009) and Chinese Academy of Sciences.

† Permanent address: Department of Physics, National University of Defense Technology, Changsha 410073, China
* klgao@wipm.ac.cn

[1] T. Udem *et al.*, Nature **416**, 233 (2002).
[2] S. T. Cundiff and J. Ye, Rev. Mod. Phys. **75**, 325 (2003).
[3] G. K. Campbell *et al.*, Metrologia **45** 539 (2008).
[4] T.Akatsukaet et al., Nature Physics **4**,954 (2008).
[5] H. S. Margolis *et al.*, Science **306,** 1355 (2004).
[6] C. Tamm et al., IEEE Trans. Instrum. Meas. **56**, 601 (2007).
[7] T. Rosenband *et al.*, Science **319,** 1808 (2008).
[8] M. Chwalla *et al.*, Phys. Rev. Lett. **102**,023002 (2009).
[9] J. E. Stalnaker *et al.,* Appl. Phys. B **89**, 167 (2007).
[10] F.-L. Hong *et al,*. Optics Express **13**, 5253 (2005).
[11] Recommendation 2(c2-2009)-(CIPM).
[12] P. A. Barton, C. J. S. Donald, D.M. Lucas, D. A. Stevens, A.M. Steane, and D. N. Stacey, Phys. Rev. A **62**,

032503 (2000).
[13] K. Matsubara *et al.*, Appl. Phys. Express **1,** 067011 (2008).
[14] Y. Huang *et al.*, Phys. Rev. A, **84,** 053841(2011)
[15] H. L. Shu *et al.*, Chin. Phys. Lett. **24,** 1217 (2007).
[16]  G. Bin *et al.*, Front. Phys. China **4,** 144 (2009).
[17] Q. Liu *et al.,* Chin. Phys. Lett. **28** ,013201 (2011).
[18]  H. Guan *et al.*, Opt. Commum. **284,** 217 (2011).
[19] G. Barwood *et al*., IEEE Trans. Instrum. Meas. **50,** 543(2001).
[20] J. E. Bernard *et al*., Phys. Rev. Lett. **82,** 3228 (1999).
[21] D. J. Berkeland *et al.*, J. Appl. Phys. **83**, 5025 (1998).
[22] A. A. Madej *et al.*, Phys. Rev. A **70**, 012507 (2004).
[23] W. M. Itano, J. Res. NIST **105**, 829 (2000).
[24] M.S.Safronoav *et al.,* Phys. Rev. A **83**, 012503 (2011).
[25] W. Lewandowski, *et al.*, Proc. IEEE **87**, 163 (1999).
[26] J. Ray, *et al.*, Metrologia **42,** 215 (2005).
[27] Bureau International des Poids et Mesures (BIPM), Circular T, May&June 2011, http://www1.bipm.org/en/scientific/tai/time_ftp.html.